\documentclass{aastex631}

\usepackage{amsmath}
\usepackage{multirow}
\usepackage{graphicx}
\usepackage{float}
\usepackage{hyperref}
\hypersetup{unicode=true}

\shorttitle{Deep Learning Diffusion Model for Cassini Background Estimation}
\shortauthors{Chen et al.}

\begin{document}
\nolinenumbers
\title{
Background Intensity Estimation for Cassini-ISS Image Using Deep Learning–Based Diffusion Model
}


\author{Yongxin Chen}
\affiliation{
Department of Computer Science, Jinan University,
Guangzhou 510632, P. R. China
}
\email{}

\author[0000-0003-4086-9678]{Qingfeng Zhang}
\affiliation{
Department of Computer Science, Jinan University,
Guangzhou 510632, P. R. China
}
\affiliation{
Sino-French Joint Laboratory for Astrometry, Dynamics and Space Science, Jinan University, Guangzhou 510632, P. R. China
}
\email{}

\author{Tianle Zhou}
\affiliation{
China Unicom (Zhejiang) Industrial Internet Co., Ltd. , Hangzhou 310000, P. R. China
}
\email{}

\author[0000-0001-7747-5707]{Kai Tang}
\affiliation{
 Shanghai Astronomical Observatory, Chinese Academy of Sciences, Shanghai 200030, P. R. China
}
\email{}

\correspondingauthor{Qingfeng Zhang}
\email{tqfz@jnu.edu.cn}


\begin{abstract}
Accurate background intensity estimation is crucial for precise astrometric measurements in astronomical imaging, particularly in complex scenarios such as those encountered in Cassini Imaging Science Subsystem (ISS) observations of Saturn's ring system. Traditional methods, like polynomial fitting and  statistical method, often fail in non-uniform conditions, such as those caused by Saturn’s rings or scattered light, due to mismatched assumptions and reliance on prior knowledge. This results in biased estimates and poor generalizability. We propose a deep learning framework based on Denoising Diffusion Probabilistic Model (DDPM) to address these challenges. By learning noise patterns and iteratively reconstructing backgrounds, DDPM improve background intensity estimation in ring\mbox{-}gap regions of ISS images by up to 62\% relative to polynomial fitting. Additionally,  When applied to centroiding of unresolved satellites in ring-gap, DDPM-based background estimation enhances positional precision by about 14\% in the line direction and 15\% in the sample direction.  The framework autonomously captures spatial correlations, requires no manual parameter tuning, and generalizes across diverse backgrounds. These characteristics make it a promising, assumption-light solution for background estimation tasks in astrometry, photometry, and source detection, with potential applications to exoplanet transit imaging, deep-field surveys, and future missions. 

\end{abstract}


\section{Introduction} 

Accurate estimation of the background intensity in astronomical images is one of the foundational requirements in quantitative astronomy, directly affecting the reliability of source detection, photometric calibration, and astrometric measurements \citep{stetson1987daophot,bertin1996}. In astrometry particularly, small systematic errors in the background model propagate to centroid determinations, biasing positions and degrading orbit solutions and dynamical inferences.

Classical approaches to background estimation (BE), such as local mesh-based sampling (e.g., sigma-clipped mean, median, or mode) and low-order polynomial surface fitting, perform well for uniform or slowly varying backgrounds \citep{stetson1987daophot,bertin1996}. However, real astronomical images often contain complex, spatially variable backgrounds arising from scattered light, instrumental artifacts, or diffuse celestial emission. Polynomial fitting and interpolation methods have been introduced to model such variations \citep{Traficante2015A&A,popowicz2015method}, yet they require a priori choices of fitting domain, masking strategy, and model order/basis. These subjective choices can under- or over-fit the true background, leaving  biased residuals that corrupt downstream photometry and astrometry.

The Cassini Imaging Science Subsystem (ISS) provides a salient example. ISS images of Saturn’s inner moons obtained near the ring plane or planetary limb exhibit strong gradients and structured stray-light patterns arising from sunlight scattered by the rings and limb-darkened planetary glare. Such non-uniform backgrounds make accurate BE challenging and can induce systematic centroid shifts \citep{zhi2024astrometry}, thereby undermining orbit determination and dynamical studies.

Recent advances in deep learning (DL) have yielded strong results across astronomical image analysis \citep{hausen2020morpheus,2022AJ....164...49D,2024ApJ...972....7R,Jia2025A&A...699A..36J,2025ApJS..279...36Z}. However, how to use DL specifically for BE remains less explored. We cast BE as a conditional completion problem: given observed context pixels, infer the latent background under source masks. This formulation naturally suggests the use of image inpainting models that condition on surrounding context. Within this family, modern inpainting methods reconstruct masked regions using learned context priors, including GAN-based approaches  \cite{zeng2022aggregated}, transformer variants \citep{quan2022image, WanICCV2021ICT, LiCVPR2022MAT}, and diffusion models \citep{sohl2015deep, saharia2022palette}.  While these models excel at synthesizing structure consistent with the neighborhood, astronomical BE demands radiometric fidelity at the pixel level rather than merely perceptual plausibility. 

In this study, inspired by recent inpainting methodologies \citep{Wang2023A, Xiang2023Deep, Quan2024},  we propose a DL-based BE approach built on  Denoising Diffusion Probabilistic Model (DDPM) \cite{ho2020denoising}. We train our model on Cassini ISS image patches with source masks and apply the model to reconstruct backgrounds in the presence of ring-induced stray light. As shown in the experiments, the method substantially improves BE in structured environments (e.g., ring gaps) while remaining competitive in smooth backgrounds, and these gains translate into more precise centroiding of unresolved satellites.

The paper is organized as follows. Section~\ref{sec:MethodData} describes the methodology and dataset. Section~\ref{sec:experiments} presents comparative experiments and astrometric evaluations. Section~\ref{sec:discussion} discusses implications and limitations, and Section~\ref{sec:conclusion} concludes.

\section{Method and Data} \label{sec:MethodData}

\subsection{Framework} \label{sec:Framework}

We developed a deep learning framework for astronomical BE tailored to Cassini ISS images. The basic idea is to mask regions occupied by celestial objects and reconstruct the underlying background using a neural generative model. In this way, the BE problem is reformulated as an image inpainting task, but with a strict requirement: the recovered intensities must faithfully represent the true astronomical signal rather than only producing visually plausible results. 

\begin{figure*}[htbp]
    \centering
    \includegraphics[width=0.5\textwidth]{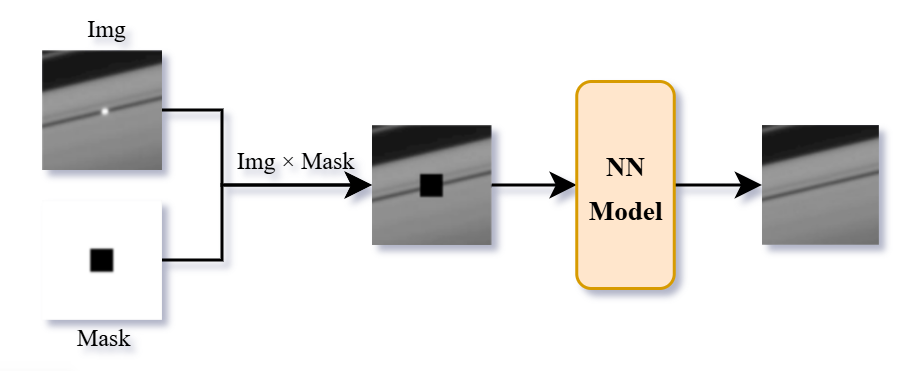}
    \caption{Celestial Background Estimation Framework Based NN Model }
    \label{fig1}
\end{figure*}

The framework has 2 stages: (1) Preprocessing: radiometric conversion, patch extraction, and source masking; (2) Background Estimation: feed masked images into a neural network trained to reconstruct missing regions using contextual information. The overall pipeline is illustrated in Figure \ref{fig1}.

 This assumption-light approach eliminates predefined model forms, making it applicable to diverse background structures. Potential neural network candidates include GAN-based inpainting models, transformer-based architectures,  diffusion models  and other generative approaches. In this study, After comparative testing, we found that DDPMs can provide the robust and accurate background reconstructions for Cassini ISS data, and we adopt them as the backbone of our implementation.

\subsection{DDPM-based Background Estimation Model} \label{sec:secBEModel}
\begin{figure*}[htbp]
    \centering
    \includegraphics[width=0.6\textwidth]{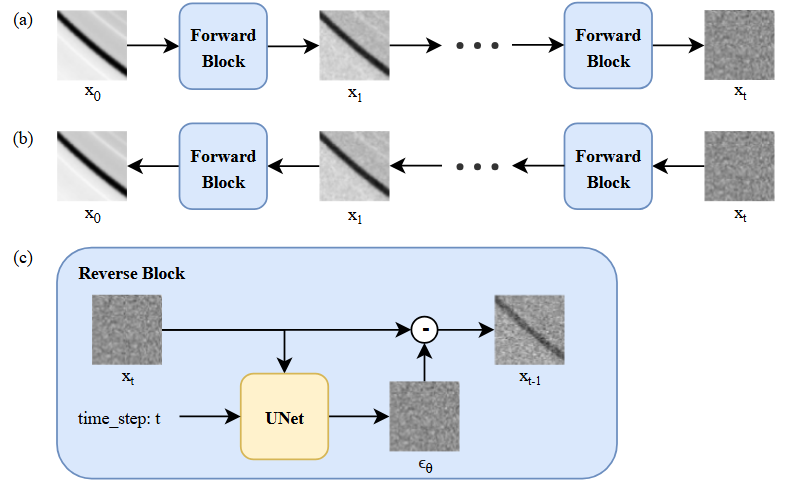}
    \caption{ \added{Illustration of the DDPM framework.} (a) Forward process; (b) Reverse process; (c) Structure of the Reverse Block in the reverse process.}
    \label{fig2}
\end{figure*}

Denoising diffusion probabilistic models are generative models that learn to reverse a noise-adding process. The process consists of two stages: a forward diffusion process that gradually adds noise to the data, and a reverse denoising process that recovers the original signal. The overall architecture is illustrated in Figure~\ref{fig2}. In the forward process (Figure~\ref{fig2}a), an original image $x_0$ is progressively perturbed over $T$ time steps, yielding a sequence of noisy representations ${x_1, x_2, \dots, x_T}$ that converge toward an isotropic Gaussian distribution.  In the reverse process (Figure~\ref{fig2}b), the model begins from a noisy input $x_t$ and iteratively denoises it to reconstruct $x_0$. Each reverse step is parameterized by a neural network, typically a U-Net \citep{ronneberger2015unet}, which predicts the added noise component $\epsilon_\theta(x_t, t)$. The Reverse Block structure (Figure~\ref{fig2}c) shows how the U-Net takes $x_t$ together with its timestep index $t$ as input, predicts the noise $\epsilon_\theta$, and subtracts it from $x_t$ via a residual connection to produce $x_{t-1}$. By repeating this operation across all time steps, the model progressively transforms Gaussian noise into a clean reconstruction of the original image.

The original DDPM \cite{ho2020denoising} is limited to generating random images from noise, which is not suitable for astronomical background estimation. Therefore, a conditional DDPM should be used to achieve our goal. We adopt an inpainting-oriented DDPM variant \cite{Lugmayr2022RePaint}. This method constrains the denoising process by incorporating known image information. During each denoising step, the model predicts the image at time $t-1$, which is then modified to ensure that the unmasked regions remain unchanged. This is mathematically represented as:

\begin{equation}
\label{eq_denoise}
\boldsymbol{x}_{t-1} 
=
\boldsymbol{m} \odot \boldsymbol{x}^{\text{unknown}}_{t-1} 
+
(\boldsymbol{1}-\boldsymbol{m}) \odot \boldsymbol{x}^{\text{known}}_{t-1}
.
\end{equation}

Here, $x_{t-1}$ is the image at time step $t-1$. \( m \) is the binary mask where 1 indicates the region to be inpainted (the celestial object and a surrounding area) and 0 indicates the known, unoccluded background.  $x^{\text{unknown}}_{t-1}$ is the model's prediction for the masked region, and $x^{\text{known}}_{t-1}$ represents the known pixels from the original image. This operation effectively combines the model's generated content within the mask with the original image content outside the mask. This ensures contextual consistency and prevents the generation of spurious artifacts, yielding a background reconstruction that is faithful to the true astronomical intensity distribution.

\subsection{Dataset} \label{Dataset Construction}
\begin{figure*}[htbp]
    \centering
    \includegraphics[width=0.5\textwidth]{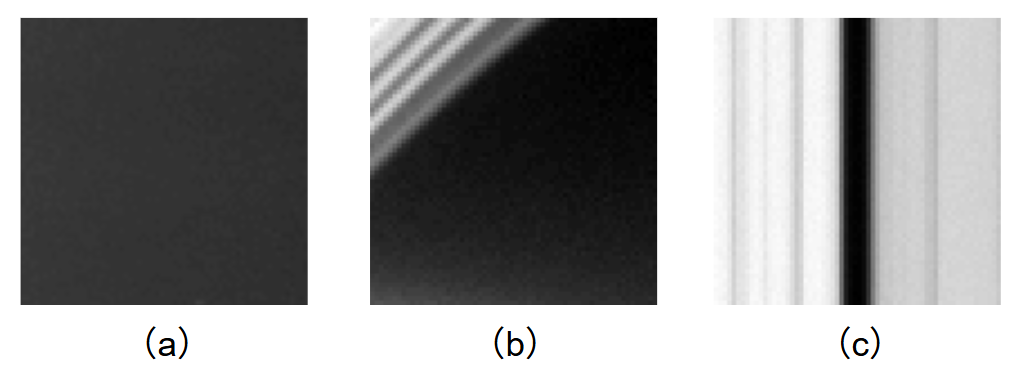}
    \caption{Example image patches illustrating the three background regimes: (a) Far-ring, (b) Near-ring, and (c) Ring-gap.}
    \label{fig3}
\end{figure*}

We constructed a dataset of $64\times64$-pixel patches extracted from Cassini ISS frames, sampling from three background regimes  defined by proximity to and illumination from Saturn's rings: ring-gap (background affected by bilateral ring-scattered light), near-ring (unilateral ring-scattered light), and far-ring (background free from ring-scattered light). To enable quantitative evaluation of BE, we deliberately selected only patches free of celestial objects. The known background intensity in central region provides a reliable ground truth, enabling a quantitative analysis of our algorithm's accuracy. Representative examples are shown in Figure \ref{fig3}.

Raw ISS images (IMG) were retrieved from the NASA archive and converted to FITS for processing, preserving headers and photometric metadata \citep{pence2010definition}. All images are 8-bit, and training and inference were performed in the same 8-bit quantization. During training, we applied standard geometric data augmentation—random $90^\circ$ rotations and horizontal/vertical flips—while avoiding photometric augmentations to preserve radiometric fidelity.

A $64\times64$ patch offers sufficient local context for BE while remaining computationally efficient.  The central $13\times13$ inpainting mask is chosen to comfortably cover unresolved point sources in ISS data (e.g., with CL1/CL2 filters the typical FWHM is $\sim\!1.3$\,pixels), while leaving a generous context margin. These choices are not claimed to be uniquely optimal; rather, they form a practical default that balances context and cost.

To avoid leakage, patches from a given raw frame are assigned to one subset—training, validation, or test—but never to multiple splits. The final dataset comprises 15,000 patches. Table~\ref{tab:tab1} summarizes the composition and split; ring-gap regions were deliberately oversampled to reflect their higher background complexity.

\begin{deluxetable*}{lccc}
\tablecaption{Composition of the training and test datasets. 
Ring-gap regions were oversampled to reflect their complex background structures.\label{tab:tab1}}
\tablewidth{0pt}
\tablehead{
\colhead{Location} & \colhead{Count (Train/Validation/Test)} & \colhead{Patch Size}  & \colhead{Gray Level} 
}
\startdata
Ring Gap   & 6000 / 500 / 500  & $64\times64$  & 8-bit \\
Near Ring  & 3000 / 500 / 500   & $64\times64$  & 8-bit \\
Far Ring   & 3000 / 500 / 500   & $64\times64$  & 8-bit  \\
\enddata
\end{deluxetable*}

\subsection{Model Training}\label{sec:training}
As outlined above, a DDPM comprises a fixed forward noising process and a learned reverse denoising process. The forward process incrementally corrupts a clean patch $x_0$ with Gaussian noise according to a prescribed schedule $\{\beta_t\}_{t=1}^{T}$ and contains no trainable parameters. The reverse process is parameterized by a neural network that predicts the standard normal noise $\epsilon_\theta(x_t,t)$ at each step, enabling iterative denoising to recover the background. In line with prior work, we adopt the perception\mbox{-}prioritized (P2) training strategy of \citet{choi2022perception} building on \citet{Dhariwal2021OpenAIGuidedDiffusion}. Our denoiser is a U\mbox{-}Net encoder--decoder equipped with multi\mbox{-}head self\mbox{-}attention (MHSA) at $1/16$ of the input spatial resolution, which captures long\mbox{-}range spatial dependencies at modest computational cost. Images are partitioned into three
regimes by background complexity---\emph{ring\mbox{-}gap}, \emph{near\mbox{-}ring} and
\emph{far\mbox{-}ring}. We use stratified train/validation/test splits and form
stratified mini\mbox{-}batches so that each regime is sampled regularly during
optimization.

Let $t\in\{1,\dots,T\}$ be the diffusion step; we use $T=1000$ during training. Following the perception\mbox{-}prioritized strategy, the objective jointly learns the noise predictor and the reverse\mbox{-}process variance via a hybrid loss

\begin{equation}
\label{eq:hybrid-obj}
\mathcal{L}_{\mathrm{hybrid}}
=
\mathcal{L}_{\mathrm{simple}}^{(\mathrm{P2})}
+
c\,\mathcal{L}_{\mathrm{VLB}},
\qquad \textnormal{with } c = 10^{-3}.
\end{equation}

where $\mathcal{L}_{\mathrm{simple}}^{\text{(P2)}}$ is a noise\mbox{-}prediction loss reweighted by the signal\mbox{-}to\mbox{-}noise ratio (P2 weighting), and $\mathcal{L}_{\mathrm{VLB}}$ is the variational lower\mbox{-}bound term. For details, refer to \cite{choi2022perception} . 
We use a linear noise schedule $\{\beta_t\}_{t=1}^T$; training draws $t$ uniformly from $\{1,\dots,T\}$. At inference we employ timestep respacing to $250$ steps with a Denoising Diffusion Implicit Model (DDIM) sampler ($\eta=0$), which yields a $\sim$4$\times$ speedup with negligible accuracy loss relative to 1000\mbox{-}step sampling.

Astronomical radiometry is preserved by operating in 8\mbox{-}bit I/O, applying only geometric augmentations (90$^\circ$ rotations; horizontal/vertical flips), and linearly normalizing patches to $[-1,1]$ without contrast stretching or histogram equalization.  Optimization uses Adam with an initial learning rate of $2\times10^{-5}$, an exponential moving average (EMA) with decay $0.999$, and global\mbox{-}norm clipping at $1.0$. Early stopping selects the checkpoint with the best validation RMSE (macro\mbox{-}averaged over the three regimes) with a patience of 10 validation evaluations. Unless otherwise specified, the mini\mbox{-}batch size is 16.

\begin{deluxetable*}{ll}
\tablecaption{Key hyperparameters used for the DDPM background predictor.\label{tab:hparams}}
\tablehead{\colhead{Setting} & \colhead{Value}}
\startdata
\sidehead{\textbf{Core diffusion}}
num\_timesteps (train)            & 1000 \\
timestep\_respacing (inference)   & 250 \\
noise\_schedule                   & linear \\
sampler                           & DDIM ($\eta=0$) \\
attention\_resolutions            & MHSA at $1/16$ scale \\
\sidehead{\textbf{Optimization \& data handling}}
optimizer                         & Adam ($\beta_1{=}0.9,\,\beta_2{=}0.999$) \\
learning\_rate                    & $2\times10^{-5}$ \\
batch\_size                       & 16 \\
EMA decay                         & 0.999 \\
gradient\_clip (global norm)      & 1.0 \\
early\_stopping                   & val RMSE (macro-avg), patience $=10$ \\
augmentations                     & $90^\circ$ rotations; H/V flips only \\
normalization \& quantization     & linear to $[-1,1]$; 8-bit I/O \\
\enddata
\end{deluxetable*}

\section{Experiments} \label{sec:experiments}
To evaluate the performance of the proposed DDPM-based BE method, we designed two experiments:  (1) a model comparison experiment, where we benchmarked our approach against traditional and representative deep learning methods; and (2) an astrometric application experiment, where we assessed the practical effectiveness of applying DDPM-based BE to ISS images for astrometric measurements. 

\subsection{Model Comparison} \label{Model_Comparison}
We conducted controlled experiments using the dataset described in Table~\ref{tab:tab1}. We randomly selected 200 images each from the ring-gap, near-ring, and far-ring categories. All selected images contain only background structures and no celestial objects. Although our final goal is to estimate backgrounds in the presence of objects, such images would not provide the true background intensities beneath the foreground, preventing direct validation. Thus, object-free images were used, ensuring that the background was fully known at every pixel.

\subsubsection*{Methods}
For each image we predicted the pixel intensities within a central $13\times13$ mask and compared the predictions against ground truth. We evaluated three approaches:

1. Traditional Methods (Polynomial Surface Fitting and Mode Estimation): Polynomial surface fitting, a classical approach for structured backgrounds, was applied to the central region (treated as a hypothetical foreground) in ring-gap and near-ring images. Unlike machine learning methods with fixed configurations ($64\times64$ patches, $13\times13$ masks), polynomial fitting required flexible choices of fitting region, background area, and polynomial order, which were optimized manually for each image. Once the optimal model was determined, background intensities for the $13\times13$ region were predicted and compared to the truth. For far-ring images with smoother backgrounds, we adopted the mode estimation method, a widely used statistical technique for sky background determination \citep{stetson1987daophot,bertin1996}, where the mode of the pixel intensity distribution (excluding the central mask) was taken as the constant background value.  

2. Aggregated COntextual-Transformation GAN (AOT-GAN) (Deep Learning Baseline):  a representative GAN-based inpainting model \cite{zeng2022aggregated}, known to outperform the methods such as DeepFill \citep{Yu_2018_CVPR,Yu_2019_ICCV} and HiFill \cite{Yi_2020_CVPR}. We adopted it as the baseline deep learning method. Following the general BE framework described in Section \ref{sec:Framework}, we specified AOT-GAN as the neural network backbone to construct an AOT-GAN–based BE model, which was trained and tested using the same dataset. After training, the model received $64\times64$ patches with a $13\times13$ central mask as input, and the predicted intensities within the masked region were directly compared to the ground truth.

3. DDPM-based BE (Ours):   The  method has already been described in Section \ref{sec:MethodData}. Similar to the AOT-GAN baseline, it follows the same BE framework but replaces the AOT-GAN with a DDPM. The trained DDPM BE model was then used to predict background intensities in the central region, and the results were evaluated against the ground truth. 

For all methods, errors were quantified using Root Mean Square Error (RMSE), bias (mean signed error), and residual Standard Deviation (STD).

\subsubsection*{Results and Analysis}
The error statistics for three methods across the three background regimes are summarized in Table~\ref{tab:table3} . The Traditional baseline delivers very low errors in the smooth far\mbox{-}ring regime but degrades markedly in structured ring\mbox{-}gap scenes, yielding large RMSE and STD. AOT\mbox{-}GAN improves over the Traditional method in ring\mbox{-}gap and near\mbox{-}ring on RMSE/STD, but its bias magnitudes remain relatively large and it underperforms the Traditional baseline in far\mbox{-}ring. In contrast, our DDPM attains the lowest RMSE and STD in ring\mbox{-}gap and near\mbox{-}ring, reducing RMSE by $\sim$62\% (11.20$\rightarrow$4.30~DN) and $\sim$54\% (4.26$\rightarrow$1.94~DN), respectively, relative to the Traditional baseline. In the nearly uniform far-ring regime, all methods achieve sub-DN errors. DDPM’s RMSE (0.85 DN) is higher than the Traditional baseline (0.38 DN) and slightly lower than AOT-GAN (0.90 DN), indicating that DDPM’s main advantage manifests in structured backgrounds.

\begin{deluxetable*}{l ccc ccc ccc}
\tabletypesize{\footnotesize}
\tablewidth{0pt}
\tablecaption{Error statistics (units: DN) across background regimes by method.\label{tab:table3}}
\tablehead{
\colhead{Method} &
\multicolumn{3}{c}{ring\mbox{-}gap} &
\multicolumn{3}{c}{near\mbox{-}ring} &
\multicolumn{3}{c}{far\mbox{-}ring} \\
\cline{2-4}\cline{5-7}\cline{8-10}
\colhead{} &
\colhead{RMSE} & \colhead{Bias} & \colhead{STD} &
\colhead{RMSE} & \colhead{Bias} & \colhead{STD} &
\colhead{RMSE} & \colhead{Bias} & \colhead{STD}
}
\startdata
Traditional   & 11.20 & -2.99 & 7.38 &  4.26 & \textbf{0.12} & 4.14 & \textbf{0.38} & \textbf{0.21} & \textbf{0.25} \\
AOT\mbox{-}GAN &  6.33 & -2.65 & 5.00 &  2.39 & 0.71         & 1.49 & 0.90         & 0.55         & 0.64 \\
DDPM (ours)   & \textbf{4.30} & \textbf{-1.33} & \textbf{2.22} & \textbf{1.94} & 0.44 & \textbf{1.30} & 0.85 & 0.47 & 0.63 \\
\enddata
\tablecomments{Bold numbers mark the best in the same column. DN = digital number.}
\end{deluxetable*}

\begin{figure*}
    \centering
    \includegraphics[width=1\textwidth]{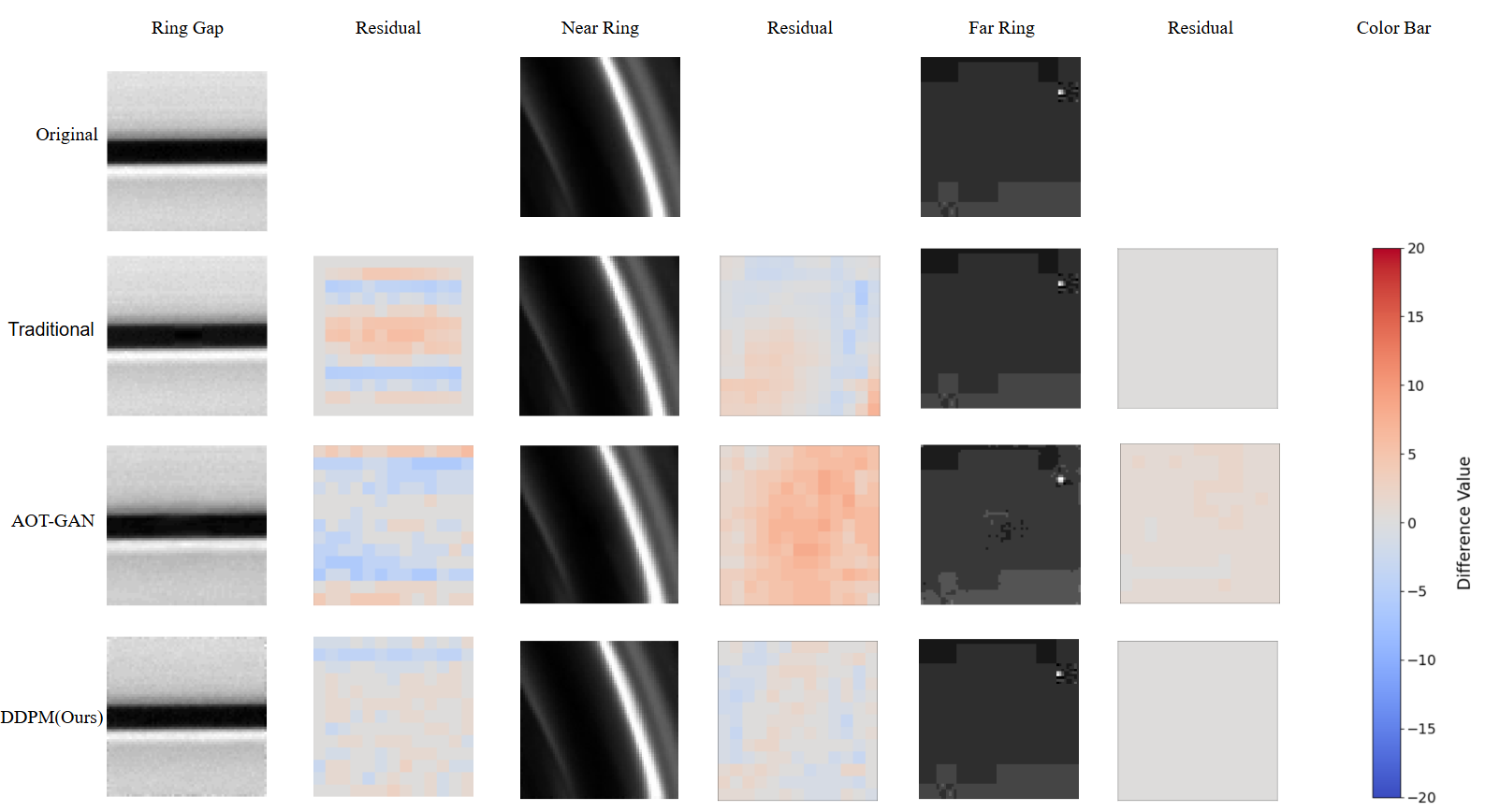}
    \caption{BE performance of traditional methods and DDPM in the absence of astronomical objects. The residual panels show two-dimensional residual maps of the central 13×13 pixel region, computed as the difference between the predicted and original images. Darker colors indicate larger residual values, with the color scale representing the magnitude of reconstruction errors.}
    \label{fig4}
\end{figure*}

Figure~\ref{fig4} provides a visual, qualitative example across the three background regimes. 
Columns 1, 3, and 5 show the original image together with the reconstruction obtained by replacing the central $13\times13$ patch with the background intensities predicted by each method; 
columns 2, 4, and 6 display residual maps over the same patch (prediction minus original) using a common color scale. 
Visually, the Traditional method leaves larger residuals in the \emph{ring\mbox{-}gap} and \emph{near\mbox{-}ring} cases, whereas DDPM yields smaller, less coherent residuals. 
In the \emph{near\mbox{-}ring} regime, the residual maps from DDPM and the Traditional method appear broadly similar, with no striking visual differences. 
These impressions are consistent with the quantitative statistics in Table~\ref{tab:table3}.

Overall, DDPM yields the lowest errors in the structured regimes—ring\mbox{-}gap and near\mbox{-}ring—surpassing both AOT\mbox{-}GAN and the Traditional baseline (polynomial surface fitting in ring\mbox{-}gap/near\mbox{-}ring; a  mode estimation method in far\mbox{-}ring). In the nearly uniform far\mbox{-}ring regime, all methods achieve sub\mbox{-}DN residuals, and the visual differences between DDPM and the Traditional mode estimation method are small; simple mode\mbox{-}based backgrounds estimator remain competitive there. These results indicate that DDPM is an effective background estimator—showing its largest gains when spatial structure is present and performance comparable to the traditional approach in smooth backgrounds.

\subsection{Application to Astrometric Centroiding}

To investigate whether improved BE translates into higher-precision centroiding of Saturn’s satellites in ISS images, we conducted an additional astrometric experiment. 100 images have been selected per category—far-ring, near-ring, and ring-gap—for a total of 300 images, each containing a  satellite. The satellites appear at different projected geometric relations to the rings: some within ring gaps, some adjacent to the rings, and others far away. These configurations reflect varying levels of scattered light contamination from the rings. All satellites appear as unresolved point sources.

Astrometric reduction of Cassini ISS images typically involves reference star detection, pointing correction, centroiding, geometric distortion correction, phase correction, and transformation between the ICRS and image coordinates \citep{Cooper2018A&A,zhang2022complementary}. Reference stars were taken from the Gaia DR3 catalog \citep{Gaia2023A&A...674A...1G,Gaia2023A&A...674A..28F}. The geometric distortion model follows Owen’s formulation \citep{2003Owen}, while phase correction is based on the theoretical treatment of \citet{Lindegren1977A&A....57...55L} and \citet{Cooper2006Icar..181..223C}. The  reduction workflow was performed using the Caviar package \citep{Cooper2018A&A}, within which we developed customized modules for background estimation and moment-based centroiding \citep{zhang2021comparison}. Because accurate background determination is critical for centroiding, we compared two BE strategies: a traditional approach and a DDPM-based approach, with all other reduction steps kept identical.

Traditional BE followed the same setup as in Section~3.1: polynomial fitting (with manually optimized parameters) for ring-gap and near-ring cases, and the mode estimation method for far-ring cases. For DDPM-based BE, we first obtained a rough estimate of the satellite center (by visual inspection), then cropped a $64\times64$ patch centered on this position, masking the $13\times13$ central region as input to the trained DDPM model.

For each image, satellite positions derived using different background estimation (BE) methods were compared against the corresponding JPL ephemerides. Because the dataset comprises several satellites—Pan, Atlas, and Anthe—we adopted the latest appropriate JPL solutions, namely SAT291 for Pan, SAT393 for Atlas, and SAT415 for Anthe. Table~\ref{tab:combined_centering} summarizes the residuals across image categories. Results show that DDPM-based BE improves centroiding in structured environments: for ring-gap and near-ring cases, residuals were reduced by about 14–15\% in the sample and line direction. In far-ring regions with smooth backgrounds, DDPM achieved accuracy comparable to traditional methods, with differences remaining within the expected measurement uncertainty.  

Overall, these results demonstrate that DDPM-based BE offers a unified and robust strategy for background estimation: it can improves centroiding accuracy in complex environments while remaining competitive in simpler ones.

\begin{deluxetable*}{lcccccc}
\tablecaption{Astrometric residuals of satellite centroids across ring-relative regimes. 
Residuals are reported  along the Cassini image axes (Sample, Line). \label{tab:combined_centering}}
\tablewidth{0pt}
\tablehead{
\colhead{Regime} & \colhead{Method} & \multicolumn{2}{c}{Sample (px)} & \multicolumn{2}{c}{Line (px)} \\
\cline{3-4} \cline{5-6}
& & \colhead{Mean} & \colhead{STD} & \colhead{Mean} & \colhead{STD}
}
\startdata
Ring Gap  & Traditional & -0.08 & 0.41 & -0.04 & 0.51 \\
          & Ours        & -0.08 & 0.35 &  0.02 & 0.44 \\
Near Ring & Traditional & -0.01 & 0.20 & -0.04 & 0.14 \\
          & Ours        & -0.02 & 0.17 & -0.05 & 0.12 \\
Far Ring  & Traditional & -0.04 & 0.09 & -0.02 & 0.11 \\
          & Ours        & -0.06 & 0.13 & -0.03 & 0.14 \\
\enddata
\end{deluxetable*}

\section{Discussion} \label{sec:discussion}

Our experiments show that the proposed DDPM-based background estimator improves BE accuracy in structured backgrounds without hand-tuned functional assumptions. Gains are largest in ring-gap and near-ring regimes, where stray-light structure challenges classical smooth models. Below we discuss implications, limitations, and prospects.

Compared with traditional polynomial fitting and AOT-GAN inpainting, DDPM achieves lower errors in \emph{ring\mbox{-}gap} and \emph{near\mbox{-}ring}. Low-order polynomial surfaces cannot represent steep, spatially localized gradients and thus leave coherent residuals; GANs tend to suppress small-scale variance because adversarial/perceptual objectives favor visual plausibility over radiometric fidelity. By contrast, DDPM reconstructs intensity distributions iteratively and probabilistically, capturing fine-scale spatial correlations without an explicit functional form. This explains the large gains in structured regimes.

Accurate BE matters because residual structure in the background directly propagates into centroiding. Our centering experiments indicate an $\sim$14--15\% reduction in positional scatter for satellites observed within or near the rings, with small and calibratable changes in mean bias. Although the absolute pixel gains are modest per frame, they are operationally meaningful: reduced scatter improves weighted combinations across sequences and mitigates background-induced shifts that can otherwise bias orbital solutions.

We adopted $64{\times}64$ patches and a $13{\times}13$ masked region to balance context and difficulty: the patch must be large enough to model ring structure surrounding the target, while the mask should be small enough to keep the inpainting problem well constrained yet large enough to cover the object and immediate halo. In principle, increasing the patch size (e.g., to $96{\times}96$) provides more low-frequency context and may help reduce structured residuals, although it could also introduce unrelated features and increase computational cost. Conversely, decreasing the patch risks truncating gradients that may be important for background estimation. Enlarging the mask (e.g., to $17{\times}17$) increases computational load and may also lead to higher reconstruction errors, whereas shrinking it (e.g., to $9{\times}9$) makes the task easier but could cause partial leakage of target flux into the surrounding context. We therefore regard the $64{\times}64$ / $13{\times}13$ configuration as a reasonable and conservative operating point, while a systematic ablation of these hyperparameters is deferred to future work.

There are some Limitations in our methods. First, Cassini ISS data are 8-bit, which compresses dynamic range and may limit ultimate radiometric fidelity; higher bit-depth data would likely benefit more from diffusion modeling. Second, our training is Cassini-specific; while the framework is model-free, direct application to other instruments without re-training may degrade accuracy. Third, the reported O$-$C residuals mix measurement noise with small ephemeris, pointing, and distortion-model errors; pixel-phase effects from undersampling can also limit centroid accuracy independently of BE quality.

The diffusion approach is assumption-light and data-driven, with utility wherever structured backgrounds limit precision—e.g., crowded-field astrometry, high-contrast imaging, and wide-field surveys with complex sky backgrounds. Across operational and upcoming facilities, scalable diffusion-based background estimation can be integrated into standard pipelines to complement physics-based scattering models and to reprocess archival data.

In the future, Several directions merit further development: (i) extend training to higher bit-depth and multi-band data; (ii) couple diffusion priors with physical (ring/scatter) forward models to enhance interpretability; (iii) perform targeted ablations of patch/mask sizes and sampler settings; and (iv)  design improved neural network architectures for more accurate background-intensity prediction.

\section{Conclusion} \label{sec:conclusion}
We have presented a novel framework for astronomical background estimation based on Denoising Diffusion Probabilistic Models and validated it with Cassini ISS data. Compared with traditional method (polynomial fitting or mode estimation method) and AOT-GAN inpainting, DDPM achieve substantially improved accuracy in structured environments while maintaining competitive performance in smooth regions. When applied to astrometric centroiding, the method enhances positional precision by 14–15\% for Saturnian satellites in structured background, confirming its value for quantitative astronomy.

The assumption-light, data-driven nature of our approach eliminates subjective parameter choices, adapts naturally to complex environments, and provides a reproducible and objective solution for background characterization. While limitations remain in data depth, generalizability, and interpretability, this work establishes DDPM as a promising new paradigm for astronomical image analysis, with implications extending from planetary science to exoplanet studies and deep-field surveys.

\section*{Acknowledgments}
This work has been partly supported by National Natural Science Foundation of China  (No. 12373073, No. U2031104),  Guangdong Basic and Applied Basic Research Foundation (No. 2023A1515011340), and Shanghai 2023 "Science and Technology Innovation Action Plan", Natural Science Foundation of Shanghai (No. 23ZR1473900).  This work has made use of data from the European Space Agency (ESA) mission {\it Gaia} (\url{https://www.cosmos.esa.int/gaia}), processed by the {\it Gaia} Data Processing and Analysis Consortium (DPAC, \url{https://www.cosmos.esa.int/web/gaia/dpac/consortium}). Funding for the DPAC has been provided by national institutions, in particular the institutions participating in the {\it Gaia} Multilateral Agreement.

\bibliography{main}
\bibliographystyle{aasjournal}

\end{document}